\documentclass[%
reprint,
superscriptaddress,
amsmath,amssymb,
aps,
pra,
floatfix,
]{revtex4-1}

\usepackage{graphicx}% Include figure files
\usepackage{dcolumn}% Align table columns on decimal point
\usepackage{bm}% bold math
\usepackage{subfigure}
\usepackage{amsmath}
\usepackage{mathrsfs}
\usepackage{amssymb}
\usepackage{setspace}
\usepackage{array}
\usepackage{multirow}
\usepackage{float}
\usepackage{flushend}
\usepackage{footmisc}
\usepackage[pdfstartview=FitH,
CJKbookmarks=true,
colorlinks,
linkcolor=blue,
anchorcolor=blue,
citecolor=blue,
urlcolor=blue,
]{hyperref}

\renewcommand{\footnoterule}{
	\kern -4pt  % 必须是负的
	\hrule width 0.18\linewidth height 0.6pt
	\kern 12pt %位置高度
}
\usepackage{braket}

\begin{document}

\preprint{APS/123-QED}

\title{Determination of dynamical quantum phase transition for boson systems using the Loschmidt cumulants method}
% \thanks{A footnote to the article title}%

\author{Pengju Zhao}
\email{1801110090@pku.edu.cn}
\affiliation{School of Physics, Peking University, Beijing 100871, China}

\author{Jingxin Sun}%
%\email{1501111282@pku.edu.cn}
\affiliation{School of Electronics, Peking University, Beijing 100871, China}
% \email{xiongfangyu@pku.edu.cn}

\author{Shengjie Jin}
%\email{lidp@pku.edu.cn}
\affiliation{School of Physics, Peking University, Beijing 100871, China}

\author{Zhongshu Hu}
%\email{lidp@pku.edu.cn}
\affiliation{School of Physics, Peking University, Beijing 100871, China}

\author{Dingping Li}
%\email{lidp@pku.edu.cn}
\affiliation{School of Physics, Peking University, Beijing 100871, China}

\author{Xiong-Jun Liu}
%\email{xiongjunliu@pku.edu.cn}
\affiliation{School of Physics, Peking University, Beijing 100871, China}

\author{Xuzong Chen}%
\email{xuzongchen@pku.edu.cn}
\affiliation{School of Electronics, Peking University, Beijing 100871, China}

\begin{abstract} 
	
%Achievement:
%1.get the complex Loschmidt zeros of the Loschmidt amplitude analogous to the Lee-Yang zeros.
%2,get the dynamical quantum phase transition point using remarkable small length.
%3.provide one method to measure DQPT experimentally.

We study the dynamical quantum phase transition(DQPT) of the Bose-Hubbard model utilizing recently developed Loschmidt cumulants method. We determine the complex Loschmidt zeros of the Loschmidt amplitude analogous to the Lee-Yang zeros of the thermal partition function. We obtain the DQPT critical points through identifying the crossing points with the imaginary axis. The critical points show high accuracy when compared to those obtained using the matrix product states method.
In addition, we show that how the critical points of DQPT can be determined by analyzing the energy fluctuation of the initial state, making it a valuable tool for future studies in this area.  Finally, DQPT in the extended Bose-Hubbaed model is also investigated.

%  \begin{description}
%    % \item[Usage]
%    %   Secondary publications and information retrieval purposes.
%    \item[PACS numbers]
%      67.85.-d, 75.10.Hk
%  \end{description}
\end{abstract}
\maketitle

\section{Introduction}

% 首先需要强调用中性原子的超交换相互作用模拟磁性模型有什么优势：比如说如果需要看spin wave这种的多体准粒子激发，均匀一致的光晶格会更有优势, thermal equilibrium和非平衡态(dynamics)都可以模拟
%一篇求解DQPT的论文，introduction DQPT，非平衡动力学， determination of the critical point，Bose-Hubbard model,literature review,大纲：
Nonequlibrium dynamics has draw much attention in recent years \cite{mitra2018,langen2016,dziarmaga2010,will2010,wilczek2012,goldman2014,eisert2015,poon2016}. One major achievement is the investigation of DQPT. DQPTs concern the critical behavior of many-body systems that are driven out of equilibrium via sudden quenches. In fact, There are two types of DQPTs. The first one concerns the nonanalytical change of order parameter along the nonequilibrium evolution\cite{poon2016,zhang2017}. For instance, in the reference \cite{poon2016}, it was observed that the evolution dynamics of the spin order parameter undergoes a significant change when the perturbation exceeds a certain threshold value. In this study, we specifically concentrate on the second type of Dynamical Quantum Phase Transitions (DQPT) that has been proposed in recent years \cite{heyl2013, heyl2018}.  This type of DQPT are inspired by the similarity between the Loschmidt amplitude  and the equilibrium thermal phase transitions. It is expected to capture unique features of nonequilibrium dynamics \cite{bhattacharya2017,sharma2016,heyl2017,budich2016,karrasch2017,wang2019,heyl2015,de2021}. Due to the absence of a direct connection to local order parameters, experimental verification of this particular type of DQPT remains challenging, particularly for systems with interaction.
%Beside theoretical considerations, it has also been experimentally observed \cite{jurcevic2017,flaschner2018,wang2019,guo2019}.jurcevic2017 extended dqpt in ssb system.flaschner2018dqpt in topological transition  ,wang2019 dqpt in quantum walks measure loschmidt amplitude guo2019 singlequbit ising model

\begin{figure}[htbp]
	\includegraphics[width=\linewidth]{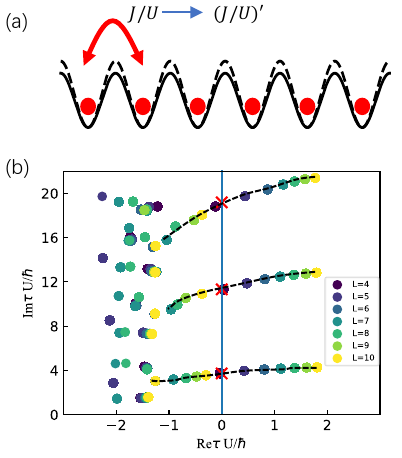}
	\caption{ Dynamical quantum phase transition. (a) A sudden quench of the hopping parameter causes a dynamical quantum phase transition in a Bose-Hubbard chain. The singularities in the rate function are associated with the zeros of the Loschmidt amplitude in the complex-time plane. These zeros form continuous lines in the thermodynamic limit, and the real critical times can be determined by identifying the crossing points with the imaginary axis. 
	}
	\label{fig:scheme}	
\end{figure}

The Bose-Hubbard model is a well-known model for investigating the many-body phases, with numerous studies conducted on both equilibrium and nonequilibrium phases. For nonequilibrium dynamics, the Kibble-Zurek mechanism \cite{huang2021,braun2015,zheng2023} and DQPTs \cite{lacki2019} have been studied for the Bose Hubbard model. A DQPT occurs in the Bose-Hubbard model when the system is quenched from the Mott-insulator phase to the superfluid phase, as shown in Fig. \ref{fig:scheme}(a).

DQPTs have been investigated by exactly solvable models \cite{heyl2013,schmitt2015,vajna2015}. However, dynamics of the interacting quantum many-body systems have been hard. Traditionally, tensor network methods are used. Recently, a new method \cite{peotta2021,brange2022} has been proposed to determine the dynamical critical points. Different from determining the critical points by identifying the nonanalytics points on the real time, the method determining zeros of the dynamical Loschmidt amplitude analogous to the Lee-Yang zeros\cite{yang1952,lee1952,brange2023} of the thermal phase transition. The critical points are determined from the crossing points of the thermodynamics lines of zeros with the imaginary axis, as shown in Fig. \ref{fig:scheme}(b). Moreover, this method offers a viable approach to determine the dynamical critical points by analyzing the energy fluctuations of the initial state. As a result, it becomes a promising avenue for investigating DQPTs.

Here we focus on DQPTs in the Bose-Hubbard model following a quench. In Sec. \uppercase\expandafter{\romannumeral2}, we introduce the basic ideas for determining the critical points using Loschmit cumulants, as illustrated in  \cite{peotta2021}. In Sec. \uppercase\expandafter{\romannumeral3}, we investigate dynamical phase transitions in the Bose-Hubbard model and accurately determine the critical points.
In Sec. \uppercase\expandafter{\romannumeral4}, we consider the nearest-neighbor interaction, and investigate quenches across topologically distinct phases. In Sec. \uppercase\expandafter{\romannumeral5}, we show that how the critical points of DQPT can be determined by analyzing the energy fluctuation of the initial state. In Sec. \uppercase\expandafter{\romannumeral6}, we present our conclusions and offer an outlook on potential directions for future advancements.

\section{Dynamical phase transition and Loschmidt cumulants}
In Reference \cite{heyl2013}, it was initially noted that a formal similarity exists between the canonical partition function
\begin{align}
	Z(\beta)=Tr e^{-\beta \hat{H}}
\end{align}
and the Loschmidt amplitude $G(t)$
\begin{align}
	G(t)=\langle \psi_0|e^{-i\hat{H}t}|\psi_0 \rangle.
\end{align}
Here, we consider the Loschmidt amplitude on the complex plane
\begin{align}
	Z(\tau)=\langle \psi_0|e^{-\tau \hat{H}}|\psi_0 \rangle.
\end{align}
According to the factorization theorem \cite{george2013}, the Loschmidt amplitude can be witten as 
\begin{align}
	Z(z)=e^{\alpha \tau} \prod_k (1-\frac{\tau}{\tau_k})
	\label{factorization}
\end{align}
where $\alpha$ is a constant, and $\tau_k$ are complex zeros of the Loschmidt amplitude. In the thermodynamics limit, the complex zeros form lines or areas. When such a line intersects or a boundary of a region touches the imaginary axis, a DQPT takes place.

Also, when a DQPT occurs, the Loschmidt rate function
\begin{align}
	\lambda(t)=-\frac{1}{L}\ln|Z(it)|^2
\end{align}
develops nonanalytic behavior.

The approach used to determine the dynamical critical points involves obtaining the line of zeros in the thermodynamic limit. As illustrated in \cite{peotta2021}, the Loschmidt cumulants and the Loschmidt moments are defined as:
\begin{align}
	&\langle \langle \hat{H}^n \rangle \rangle_{\tau}=(-1)^n \partial^n_{\tau} \ln Z(\tau)  \\ 
	\langle \hat{H}^n \rangle =&(-1)^n\frac{\partial^n_{\tau} Z(\tau)}{Z(\tau)}=\frac{\langle \psi_0|\hat{H}^n e^{-\hat{H}\tau}|\psi_0 \rangle}{\langle \psi_0|e^{-\hat{H}\tau}|\psi_0 \rangle} \label{moments}
\end{align}
At $\tau=0$, the Loschmidt moments reduce to the ordinary moments (\ref{moments}) of the postquench Hamiltonian with respect to the initial state as $\langle \hat{H}^n\rangle =\langle \Psi_0| \hat{H}^n| \Psi_0\rangle$.
%The Loschmidt cumulants and Loschmidt moments are related by the standard recursive formula.
%\begin{align}
%	\langle \langle H^n \rangle \rangle_{\tau}=\langle H^n \rangle-\sum^{n-1}_{m=1} C_{m-1}^{n-1} \langle \langle H^m \rangle \rangle_{\tau}\langle H^{n-m} \rangle
%\end{align}

The Loschmidt cumulants are related to the Loschmidt zeros through
\begin{align}
	\langle \langle \hat{H}^n \rangle \rangle_{\tau}=(-1)^{n-1} (n-1)!\sum_k \frac{1}{(\tau_k-\tau)^n}
	\label{zeros}
\end{align}

According to this expression, the Loschmidt cumulants are primarily influenced by the zeros that are closest to the base point $\tau$. The contribution of each zero to the cumulants decreases rapidly with its inverse distance to the power of the cumulant order $n$. By computing $2m$ high-order Loschmidt cumulants, it becomes possible to determine the $m$ closest zeros to the base point.

We extract the 7 zeros closest to the movable basepoint using Loschmidt cumulants of order n=9 to n=22.
In our time evolution of the wavefunction, we select the Krylov subspace dimension to be $N_{\text{vec}} = 8$ and the time step for evolution to be $\delta \tau = 0.01$.

%Finally,the coeficients $d_k^{(n)}$ can be used to justify the reliability of the numerical results.$d_k^{(n=0,1,...,m-1)}$ satisfy the following equation.
%\begin{align}
%	\left(\begin{matrix}		
%1&1&...&1&1&\\	
%	\lambda_0&\lambda_1&...&\lambda_{m-2}&\lambda_{m-1}&  \\
%	\vdots & \vdots &\ddots& \vdots &\vdots  \\
%	\lambda_0^{m-2}&\lambda_1^{m-2}&...&\lambda_{m-2}^{m-2}&\lambda_{m-1}^{m-2} \\
%     \lambda_0^{m-1}&\lambda_1^{m-1}&...&\lambda_{m-2}^{m-1}&\lambda_{m-1}^{m-1}    
% \end{matrix}\right)     
%   	\left(\begin{matrix}		
%  d_0^{(n)}\lambda_0^n \\d_1^{(n)}\lambda_1^n\\ \vdots \\ d_{m-2}^{(n)}\lambda_{m-2}^n \\d_{m-1}^{(n)}\lambda_{m-1}^n	
%    \end{matrix}\right)     
%=\left(\begin{matrix} \kappa_n \\\kappa_{n+1} \\ \vdots \\ \kappa_{n+m-2}\\\kappa_{n+m-1}   \end{matrix}\right)
%\end{align}

%For boson systems,atoms can occupy the same site. We choose the occupation number up to 3, the propability of atoms populated higher than 3 is less than 0.02\%.  

\section{the Bose-Hubbard model}

\begin{figure*}[htbp]
	\includegraphics[width=\linewidth]{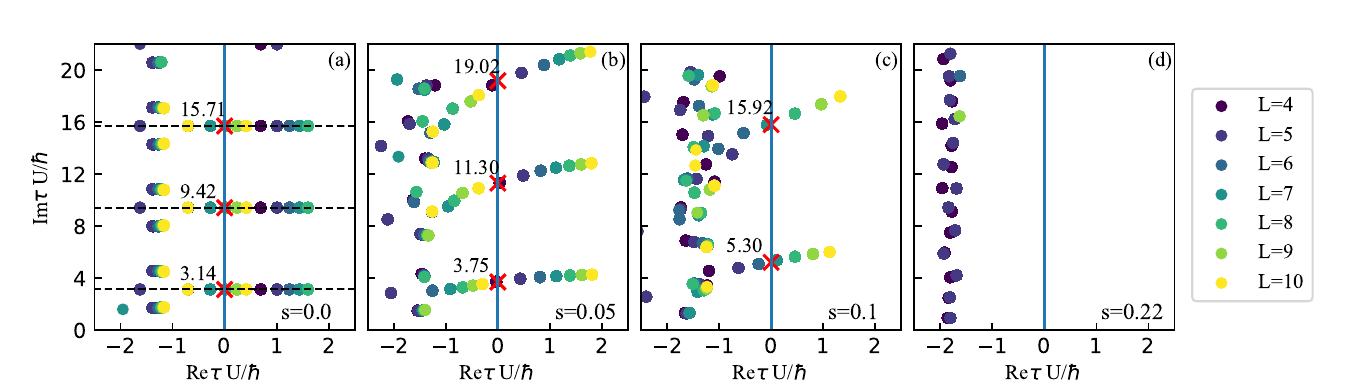}
	\caption{Determination of critical points for the Bose-Hubbard model.Complex zeros for different system sizes(L=4-10) are shown. We quench the system from the superfluid phase with initial parameter $s_0=J/U=0.36$ to the Mott phase. In panel (a), horizontal lines with Im$\tau\: U $/$\hbar=\pi,3\pi,5\pi$ are shown. The critical time $t_c$ is obtained as the intersection between the imaginary axis and the line drawn from the zeros $\tau_{-}$ with the smallest negative real part (in absolute value) to the zero $\tau_{+}$ with the smallest positive real part. The critical time obtained using MPS method (red cross) with length of 120 sites are drawn for comparison. (b)-(d)Same as in panel (a) but for different parameter values: $s=0.05$ in panel (b), $s=0.1$ in panel (c), and $s=0.22$ in panel (d).
	}
	\label{fig:0.36}	
\end{figure*}

In this study, we explore DQPTs in the one-dimensional Bose-Hubbard model, characterized by the Hamiltonian:

\begin{align}
	\hat{H} = -J \sum_i (a_{i}^{\dagger} a_{i+1} + a_{i+1}^{\dagger} a_{i}) + \frac{U}{2} \sum_{i} n_{i}(n_{i}-1) \label{eqn:bhm}
\end{align}

Here $a_i^{\dagger}$ is the creation operator for a boson on site i, $a_i$ is the annihilation operator for a boson on site i, $n_i=a_i^{\dagger}a_i$ is the occupation operator for a boson on site i, $J$ denotes the hopping amplitude between nearest neighbor, $U$ denotes the onsite interaction strength. To minimize boundary effects, we impose the periodic boundary condition.

The properties of the Hamiltonian are determined by the dimensionless ratio $s=J/U$, and a phase transition occurs at the critical value of $s_c=0.297$ for unit filling, which separates the system into a Mott-insulator phase for $s<s_c$ and a superfluid phase for $s>s_c$.

We are now prepared to explore the phenomenon of Dynamical Quantum Phase Transitions (DQPT) in the Bose-Hubbard chain. We initialize the system in the superfluid ground state with a parameter value of $s_0=0.36$. The local Hilbert space is truncated to $n=3$ for an balance of efficciency and accuracy. The atom occupation number for $n>3$ is less than 0.03\% for the initial state.

We perform a quench into the Mott-insulator phase with $s<0.297$ for later times. The same quench parameters have been explored in \cite{lacki2019} using matrix product method. Here we employ the matrix product state(MPS) method for benchmark.

Figure \ref{fig:0.36} shows the complex zeros of the Loschmidt amplitude. 
 For the $J=0$ case, the system undergoes periodic evolution with period $T=\frac{2\pi\hbar}{U}$. The first critical time is $t_1=\frac{\pi}{U}$. As expected, the zeros forms a line around $t=\frac{\pi\hbar}{U},\frac{3\pi\hbar}{U},\frac{5\pi\hbar}{U}$ on the imaginary axis.
As the tunneling amplitude is incrementally increased, it can be observed in Figure \ref{fig:0.36}(b-c) that the Loschmidt zeros gradually shift the critical crossing point with the imaginary axis towards later times. Finally the thermodynamic lines of zeros no longer crosses the imaginary axis.

The critical time obtained from the crossings of the thermodynamic lines of zeros with the imaginary axis are in excellent agreement with the critical times abtained using the MPS method. To quantify the accuracy of the Loschmidt cumulants method, we determine the critical points and compare it with those obtained using MPS method. We find that for all typical cases in Fig. \ref{fig:0.36} the discrepancy is lower than 2\%. It is worth noting that these results are obtained for rather short length from L=4 to L=10. The use of such system sizes makes the approach very attractive for strongly intracting systems.

\section{The extended Bose-Hubbard model}
\begin{figure*}[htbp]
	\includegraphics[width=\linewidth]{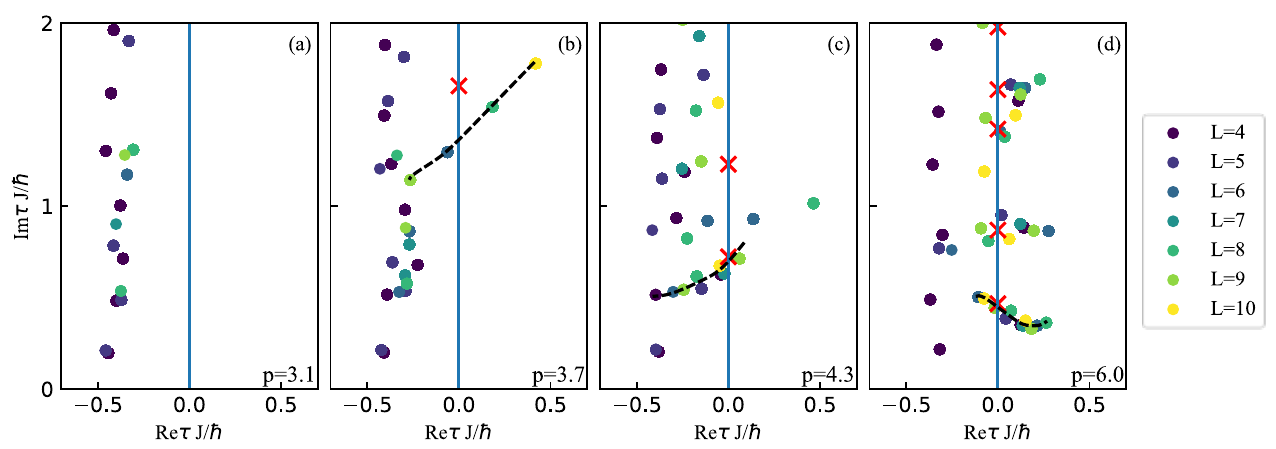}
	\caption{ Determination of critical points for the extended Bose-Hubbard model. We quench the system from the Mott insulator phase with initial parameter $U/J=5$, $p_0=V/J=1.0$. We quench the system to the final Hamiltonian of various p. Panels (a)-(d) illustrate DQPTs corresponding to various final parameters: $p=3.1$ in panel (a), $p=3.7$ in panel (b), $p=4.3$ in panel (c), and $p=6.0$ in panel (d).
	}
	\label{fig:1.0}	
\end{figure*}

We will now explore DQPTs in the extended Bose-Hubbard model, which includes nearest-neighbor interactions and exhibits rich phase diagrams such as the Haldane insulator(HI) and the charge density wave (CDW). The Hamiltonian for the extended Bose-Hubbard model is given by:
\begin{align}
	\hat{H}=&-J\sum_i (a_{i}^{\dagger} a_{i+1}+a_{i+1}^{\dagger} a_{i})+\frac{U}{2}\sum_{i} n_{i}(n_{i}-1)\\ \notag
&+	V\sum_i n_i n_{i+1}
\end{align}
Here V denotes the nearest neighbor interaction strength.  For a ratio of interaction to hopping strength of $U/J=5$, the equilibrium phase transition point is $p=V/J=2.95$ for the Mott insulator-Haldane insulator transition and $p=3.53$ for the Haldane insulator-density wave transition \cite{rossini2012}. We adopt periodic boundary conditions in this study \cite{stumper2020,ejima2014}.

We quench from the Mott phase $p_0=1.0$ to larger nearest neighbor interaction \cite{stumper2022}. The results are shown in Fig. \ref{fig:1.0}. Dynamical quantum phase transition occurs at about $p=3.5$. As we increase the ratio $p$, dynamical phase transition happens at ealier times. The discrepancy is larger near the equilibrium phase transition points and decreases for cases further away from these points. This may be due to  the finite-size effect. For further improvement, other boundary conditions may be introduced \cite{peotta2021}. 
Although our study focuses on a narrow range of parameters, it is feasible to investigate DQPTs for other parameter ranges in the extended Bose-Hubbard model as well.

%
%\begin{figure*}[tbp]
%	\includegraphics[width=\linewidth]{Figure6.png}
%	\caption{ Zeros of the Loschmidt amplitude. We quench the system from the Mott phase to the superfluid phase.The initial state is the ground state of the model with J=0.36,and unity occupation number. The blue and green dot are Loschmidt zeros calculated using Loschmidt cumulants method. The red point are the dynamical phase tansition points derived using TEBD with length of 160 sites.
%	}
%	\label{fig:1.0}	
%\end{figure*}

%
%\begin{figure*}[tbp]
%	\includegraphics[width=\linewidth]{Figure5.png}
%	\caption{ Zeros of the Loschmidt amplitude. We quench the system from the Mott phase to the superfluid phase.The initial state is the ground state of the model with J=0.36,and unity occupation number. The blue and green dot are Loschmidt zeros calculated using Loschmidt cumulants method. The red point are the dynamical phase tansition points derived using TEBD with length of 160 sites.
%	}
%	\label{fig:3.25}	
%\end{figure*}

\begin{figure*}[htbp]
	\includegraphics[width=\linewidth]{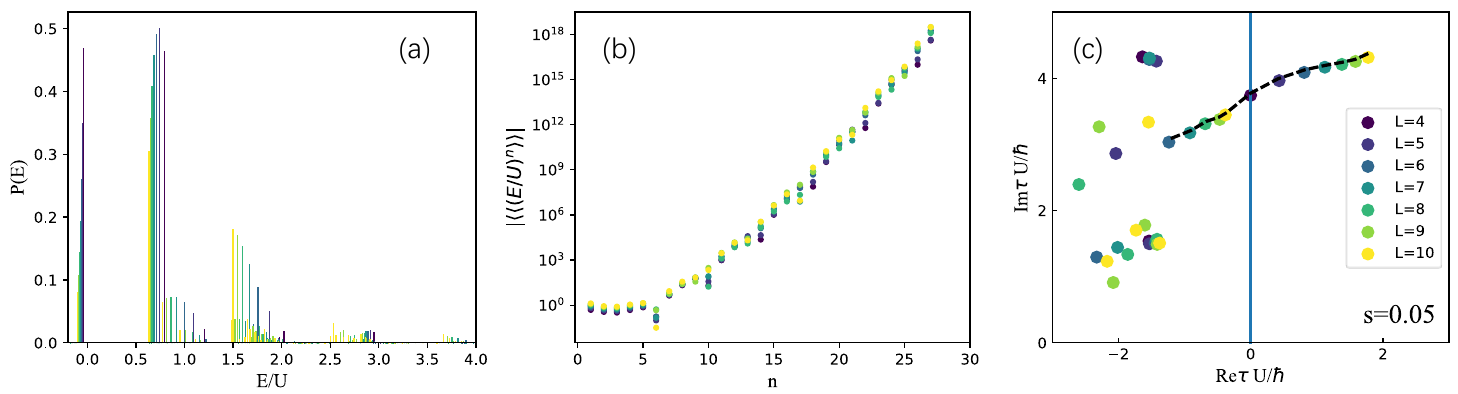}
	\caption{ Determination of the critical time from the initial energy fluctuation. We quench the system from the superfluid phase $s_0=0.36$ to the Mott insulator phase $s=0.05$. (a) The energy distribution obtained from $10^6$ energy measurements of the postquench Hamiltonian. (b)Energy cumulants determined from the energy distribution. (c)Determination of the Loschmidt zeros using cumulants of orders n=4 to n=27. The dashed lines are drawned to guide the eyes. The critical point can be identified as the intersection of the line with the imaginary axis.
	}
	\label{fig:exp}	
\end{figure*}

%	\caption{ Determination of the critical time from the initial energy fluctuation. We quench the system from the Mott phase to the superfluid phase.The initial state is at $s=0.36$. The postquech Hamiltonian is the parameter $s=0.05$. We have extracted the zeros using cumulants of orders n=. The dashed lines are drawned to guide the eyes. The critical point can be identified as the intersection of the line with the imaginary axis.
%}

\section{Experimental perspective of the Loschmidt zeros} 

%An experimental determination of Loschmidt zeros for small interacting quantum systems appears feasible with current technology.
Finally, we demonstrate the ability to predict DQPT solely through measuring the energy fluctuations in the initial state. At $\tau=0$, the Loschmidt moments reduce to the ordinary moments (\ref{moments}) of the postquench Hamiltonian with respect to the initial state as $\langle \hat{H}^n\rangle =\langle \Psi_0| \hat{H}^n| \Psi_0\rangle$. If we expand  the wavefunction with respect to the post-quench Hamiltonian $| \Psi_0\rangle=\sum_m a_m |\tilde{\Psi}_m \rangle$, we get $\langle \hat{H}^n\rangle =\sum_m P(E_m) E_m^n $, where $P(E_m)=|a_m|^2$.
Thus, by repeatedly preparing the system in the state $|\psi_0\rangle$ and measuring the energy with respect to the postquench Hamiltonian, we can construct the energy distribution and extract the corresponding moments and cumulants. Therefore, by following the same procedure presented in the method section, we can obtain the Loschmidt zeros.

In Fig. \ref{fig:exp}, we present the procedure of determination the critical points from $10^6$ energy measurements.  This predicts the critical time to be around $t\;U $/$\hbar=3.75$. This demonstrate the ability to predict the first dynamical phase transition using initial energy fluctuations. The determination of a dynamical Lee-Yang zeros was accomplished in Ref. \cite{brandner2017} through the measurement of high cumulants. 
It should be noted that obtaining precise energy measurements of a many-body quantum system is challenging. Thus, the presented method offers an approach to establish a connection between DQPTs and measurable quantities i.e energy, at least in principle.
%This scheme may be implemented in the near-future quantum simulators \cite{xu2020,guo2021}.

%Considering the high controbility\cite{gross2017,bloch2008} in cold atom systems,it is possible to measure DQPT experimentally. 
%

\section{Conclusion}
In summary, we have investigated the Loschmidt zeros associated with the dynamical quantum phase transitions in the Bose-Hubbard model and the extended Bose-Hubbard model. This analysis was conducted through the utilization of the Loschmidt cumulants method. We have determined the locations of the zeros of the Loschmidt amplitude in the complex plane of time. And by identifying the crossing point with the imaginary axis we get the dynamical phase transition point to an discrepancy lower than 2\% for the Bose-Hubbard model. Also, we show an avenue for determining the dynamical phase transition by measuring the initial state energy fluctuation.

%  As the 2D Ising model with a tranvserve magnetic field can be mapped to a 3D Ising model \cite{Suzuki1976}, it would promote the research of this problem to a large extent if we can experimentally realize an exact 2D Ising model with a tranvserve field. 

\section*{Acknowledgements}
Acknowledgements: This work is supported by the National Natural Science Foundation of China (Grants No. 11920101004,  11934002, 12174006), and the National Key Research and Development Program of China (Grant No. 2021YFA1400900, 2021YFA0718300).
%\end{acknowledgements}
%\section{Appendices}

%\vspace{-0.2cm}
%\subsection{Appendix A: }
%
%\label{appendix:A}

\bibliographystyle{apsrev4-1}
%\bibliography{ref2}

\providecommand{\noopsort}[1]{}\providecommand{\singleletter}[1]{#1}%

\end{document}